\documentclass[]{spie}  
\usepackage[]{graphicx}
\usepackage{verbatim}
\usepackage{subfigure}

\newcommand{\vv}{Vaulted Verification }
\newcommand{\vvv}{Vaulted Voice Verification }

\title{Secure voice based authentication for mobile devices:\\ Vaulted Voice Verification} 

\author{R.C. Johnson\supit{a,b}, Walter J. Scheirer\supit{a,c} and Terrance E. Boult\supit{a,b}
\skiplinehalf
\supit{a}Securics, Inc, Colorado Springs, CO, USA; \\
\supit{b}University of Colorado Colorado Springs, CO, USA; \\
\supit{c}Harvard University, Cambridge, MA, USA;\\
}

\authorinfo{Further author information: (Send correspondence to R.C. Johnson)\\R.C. Johnson and Terrance E. Boult: E-mail: \{rjohnso9 , tboult\} at uccs.edu,\\  Walter J. Scheirer: E-mail: wscheirer at fas.harvard.edu}

\begin{document} 
  \maketitle 

\begin{abstract}
As the use of biometrics becomes more wide-spread, the privacy concerns that stem from the use of biometrics are becoming more apparent. 
As the usage of mobile devices grows, so does the desire to implement biometric identification into such devices.
A large majority of mobile devices being used are mobile phones.
While work is being done to implement different types of biometrics into mobile phones, such as photo based biometrics, voice is a more natural choice.
The idea of voice as a biometric identifier has been around a long time.
One of the major concerns with using voice as an identifier is the instability of voice.
We have developed a protocol that addresses those instabilities and preserves privacy.
This paper describes a novel protocol that allows a user to authenticate using voice  on a mobile/remote device without compromising their privacy.
We first discuss the \vv protocol, which has recently been introduced in research literature, and then describe its limitations.
We then introduce a novel adaptation and extension of the \vv protocol to voice, dubbed \vvv ($V^3$).
Following that we show a performance evaluation and then conclude with a discussion of security and future work.

\end{abstract}



\section{INTRODUCTION}
\label{sec:intro}  

This paper addresses the question ``Can voice be used in a remote verification protocol that also preserves privacy?" 
This paper shows that yes it can.
We do this by borrowing techniques from the voice community and integrating them with a recently introduced protocol called \vv from the vision community.
The \vv protocol is a vision based protocol that allows for remote verification in a privacy-preserving way \cite{vv-wacv2012, vv-cvpr2012}.
The \vv protocol will be discussed further in Section \ref{sec:vv}.

Biometric security is a growing field.
In recent years, the popularity of mobile biometrics has increased tremendously with the proliferation of incorporated sensors into everyday technology, such fingerprint scanners in laptops and face recognition in mobile phones.
The increased interest in biometric security has lead to more of a focus being placed on the security of such techniques \cite{jain_nandakumar2008}.

Because the actual biometrics are limited and therefore too easily compromised, biometric security protocols must be based on templates, or models.
With this, there is much concern over how to protect these templates.
Most of the state-of-the-art template protection schemes can be grouped into one of two categories; feature transformation, or biometric cryptosystems \cite{jain_nandakumar2008}.
Schemes from the feature transformation category include things like hashing techniques that take the raw features and use a transformed version of them.
Schemes from the cryptosystems include things such as fuzzy vaults which hide the real data among the chaff.

One of the biggest issues in biometric systems is the security of enrollment and verification processes for the biometric data. 
Once a biometric sample (scan, recording, picture, etc...) has been obtained by the system the information has to be not only saved securely, but the system also has to be able to verify against it, or the resulting saved data, with a second sample in order to verity the validity of the second sample.
To solve the security issues with the processes, there are a few requirements that must be met. \cite{fuzzy_vaults2007}
These requirements are:
\begin{enumerate}
\item Ensure that the biometric can not be compromised if the server is compromised and the model is acquired by an attacker. 
\item Prevent man-in-the-middle and replay attacks.
A ``man-in-the-middle" is an attack where someone intercepts the communication from the server to the client and replaces the information on both sides.
This is similar to the ``replay" attack where an attacker saves the captured information to be used again later.
\item The system must have revocability.
If the user needs to invalidate their model current model or something happens where the model is compromised the user should be able to invalidate their current model and issue a new model.
\item Withstand blended substitution attacks. 
In the blended substitution attack, an attacker combines their template with the stored template so that both the user and the attacker can authenticate using the same template.
The system should be able to detect and prevent someone from substituting a users model with their own.
\end{enumerate}

\begin{figure}
\centering 
\subfigure[\protect\parbox{70mm}{The Problem: Multiple users trying to access server as Bob.}] {
\resizebox{80mm}{!}{\includegraphics{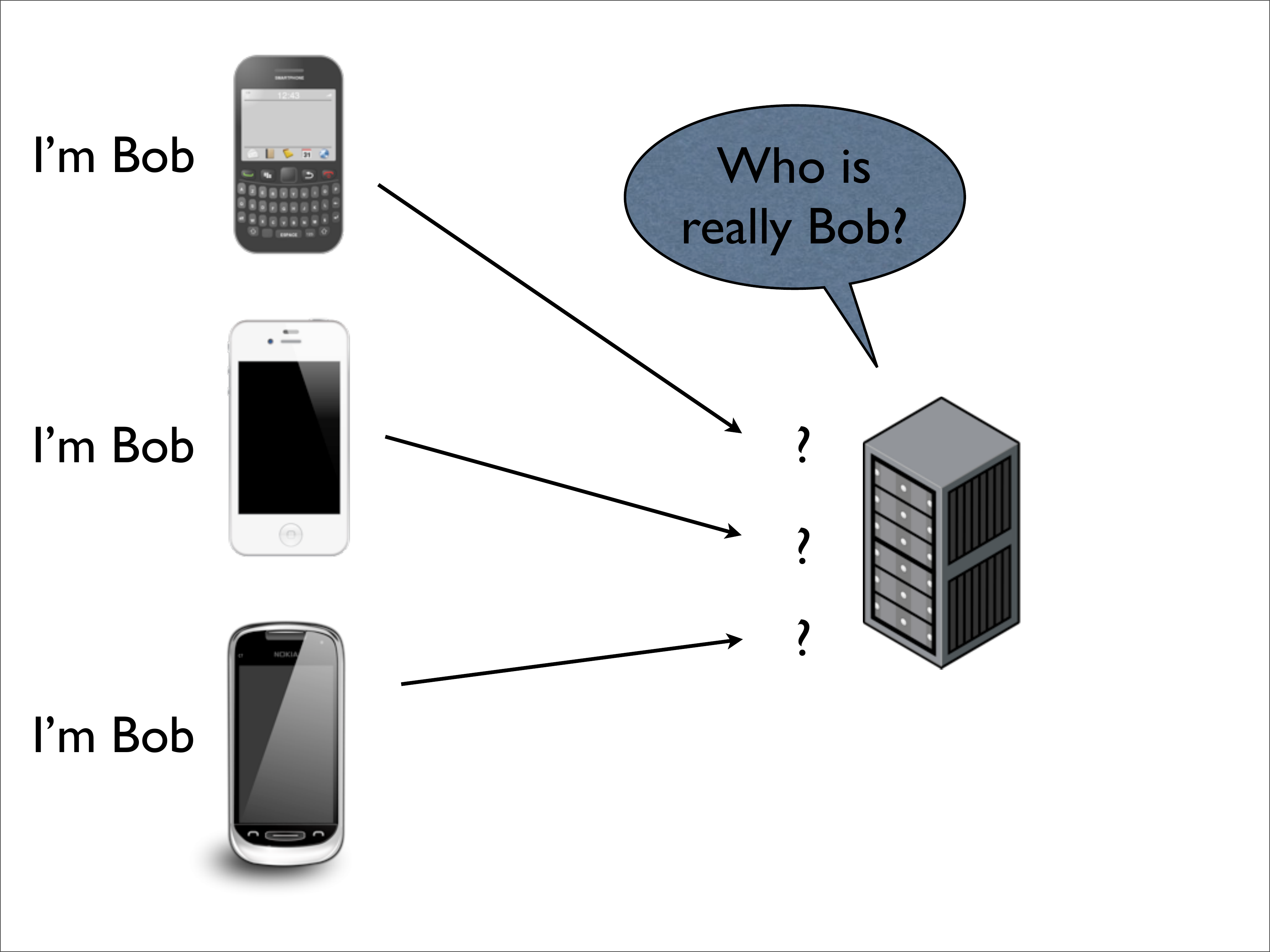}}
\label{vvv_problem}
}
\subfigure[\protect\parbox{70mm}{The Answer: Ask them to say words/phrases and match real vs imposter}]{
\resizebox{80mm}{!}{\includegraphics{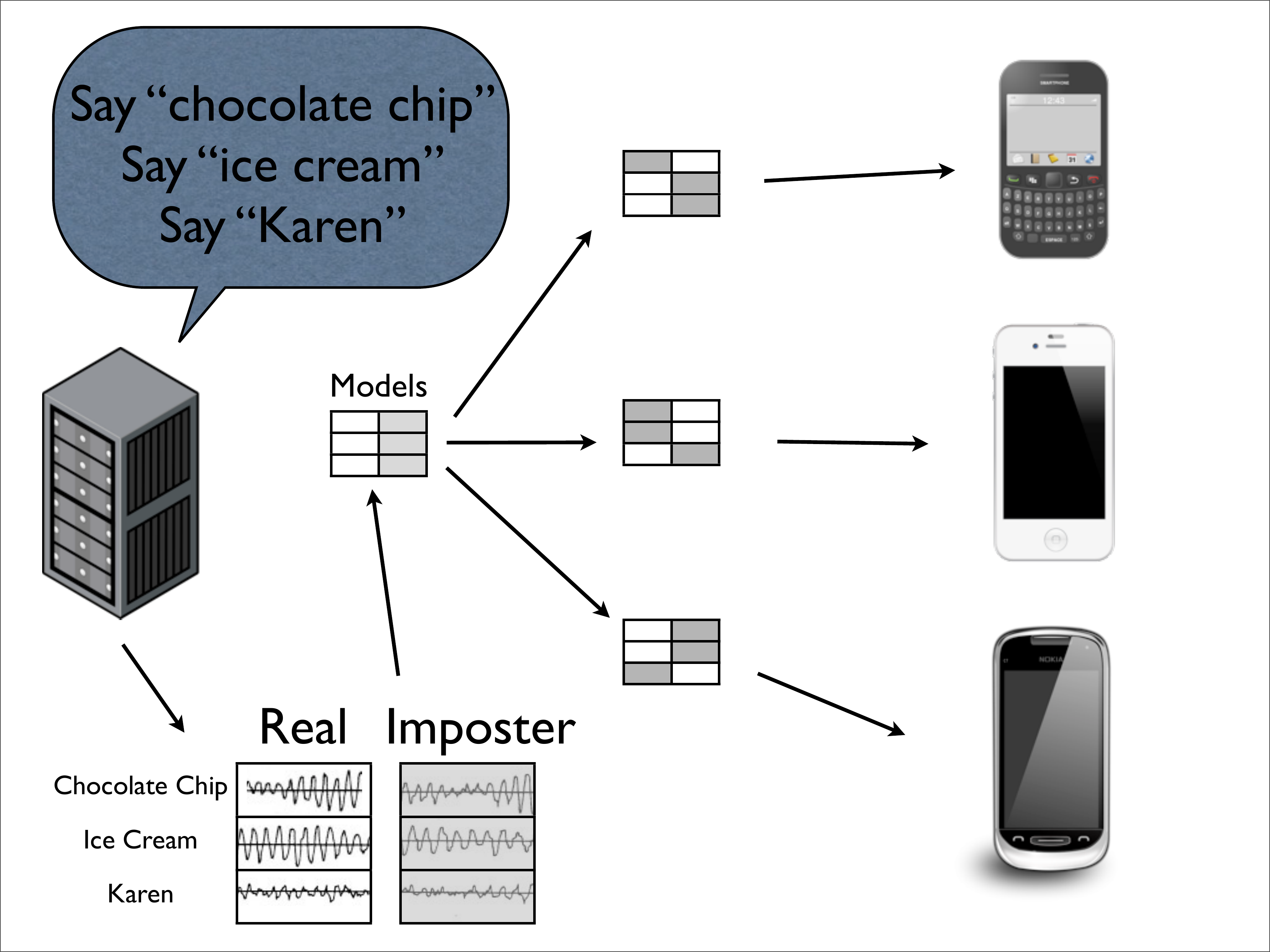}}
\label{vvv_answer}
}
\caption{Purpose of \vvv}
\label{vvv_figure}
\end{figure}

The prior work on the \vv technique addresses the previously mentioned requirements.
In the current work we are extending this to \vvv, a novel challenge-response protocol appropriate for voice based verification.
The \vvv protocol is a new challenge-response approach to authentication well suited to mobile devices. 
The problem \vvv truly solves is how can this be done in a secure way while also preserving the privacy of the user.
One of the advantages of \vvv is that security can be increased by combing speech content with the challenges.
As an example, Figure \ref{vvv_problem} shows multiple people trying to access a system while all claiming to be Bob. 
Using the novel \vvv protocol, a simplified version of which is shown in Figure \ref{vvv_answer}, the server is able to use the challenge response pairs to ensure the authenticity of the users.
This example shows the server challenging the users by having them speak three words/phrases.
The server sends pairs of models to the user, with one authentic and one chaff model for each phrase.
The user must decide which of each pair is authentic based on their response, 0 if they think the real is first and a 1 otherwise.
In Figure \ref{vvv_answer} the top user should respond 100, the middle user should respond 110, and the bottom user should respond 001.
The authentic user should have no difficulty choosing the correct bits; impostors have to guess.
There are other layers of security, the algorithm is described in greater detail in section \ref{sec:vvv}.

\section{Related Work}
\label{related}
Many techniques are used in an effort to utilize voice as a biometric identifier.
The idea of privacy and security using voice goes many years back as in 2001 when Monrose et al \cite{fabian2001} created a system for biometric key generation.
However, the system created by Monrose et al was impractical.
There has been lots of work on voice, as Jain et al \cite{jain_nandakumar2008} and Rathgeb et al \cite{rathgeb2011} discussed in their survey papers which discuss the state of the art.
Here, we are only going to discuss the most relevant papers to this work.

In the last couple of years, different groups have begun to focus on using template protection with speech using GMMs.
Teoh and Chong \cite{teoh2010} used Probabilistic Random Projections (PRP)  to protect their speaker template while doing speaker verification.
In their research, the template is hidden in a process of random projections in a subspace.
The EER obtained in their experiment range between 0\% and 8.94\% depending on the scenario.
\vvv differs from this work because it is a client-server protocol that can be used on mobile devices, and the template is not stored on the server.
\vvv also improves over the technique discussed by Teoh and Chong in that we use multiple templates instead of a single template, this improves the certainty in the final score.

In this paper, we are using MFCCs and will highlight the similarities and difference between our work and the work of others using similar structures.
For example, in the work done by Murty et al \cite{murty2006} the combination of residual phase and mel-frequency cepstral coefficients (MFCCs) effect a 10.5\% equal error rate (EER) for speaker verification in a text-independent speaker verification system with no security or protection. 

Some of  recent work by Lopresti et al \cite{lopresti2011, lopresti2012} proposes combining ``a speech biometric cryptosystem with a password" .
In this work, the biometric template is transformed, mapped, permuted, fuzzy committed, and more.
During the transformation stage of their process, a Discrete Fourier Transform (DFT) is applied to the raw audio signal to produce the feature vectors.
This is similar to, if not the same as, the process of obtaining MFCCs.
The system then performs dynamic time warping  to ``minimize distance between the reference signal and training utterance."
Next, the features are mapped to a binary string.
When features that are most stable are found, a string of them, collectively called a ``distinguishing descriptor" is used in the next step. 
To harden their template, they perturb the system by run the mapping function over and over again, each time removing one of the stable features and using this new vector as the key in the dynamic time warping.
Until the resulting distinguishing descriptor has less than or equal to half of the feature vector length, this process repeats.
At this point, the template gets transformed and permuted, as more thoroughly described in their work \cite{lopresti2011}, so that the last step is to combine their final binary string with a cryptographic key.
In the verification phase, a template, a bunch of thresholds, and the lock data are given to the user.
Then, the user's password and a new biometric submission are used to generate a new set of distinguishing descriptors, which then have the correct indexes pulled out to match the lock data.
If the user's data matches the lock data, within some tolerance, then the decoding process takes place, and the system authenticates the user.

Using their technique on the MIT mobile device speaker verification corpus they report EER using three different scenarios; neither password nor biometrics compromised, password compromised, biometrics compromised. 
The EERs reported are 0\%, 11.11\%, and 0\%, respectively for their two layer scheme. 
This shows two main things; that the technique becomes vulnerable if the password is compromised and that the dataset is not challenging enough for full evaluation of the technique.

\section{\vv}
\label{sec:vv}

As with any client-server verification process, the \vv process utilizes two main steps: enrollment and verification.
\vvv shares its base concept with the \vv protocol, so this section will provide a general overview of the \vv process \cite{vv-wacv2012, vv-cvpr2012}.

\begin{figure}
\centering
\resizebox{140mm}{!}{\includegraphics{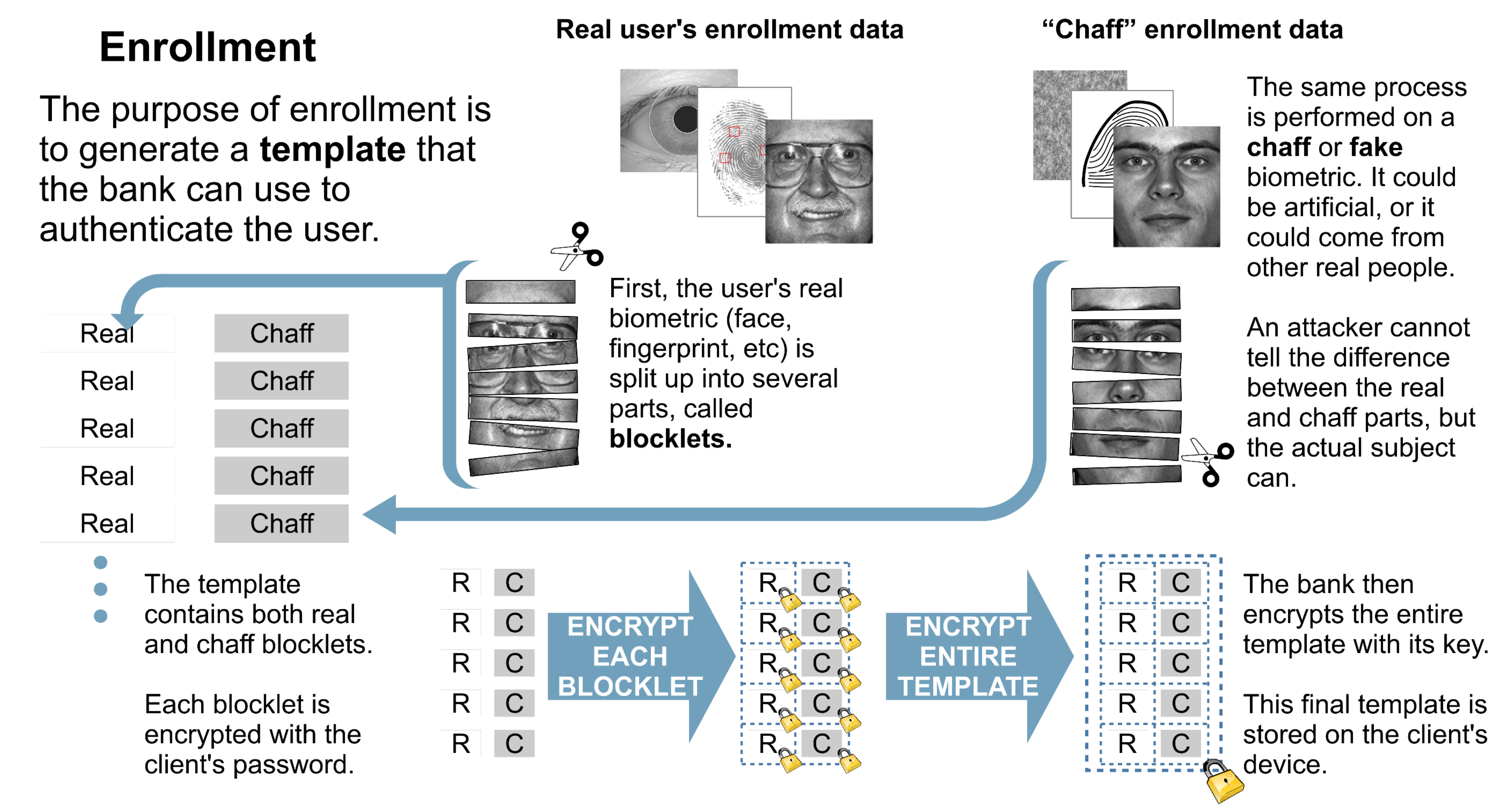}}
\caption{Visual example of \vv enrollment \cite{vv-wacv2012}.}
\label{vv_enroll}
\end{figure}

\subsection{Overview}
We first define some basic terms to assist in understanding the \vv ($V^2$) process. 
$C$ refers to the client software running on the mobile device. 
$U$ refers to a person using $C$. 
$S$ refers to the machine/device to which $U$ is trying to gain access. 
$K_S$ and $K_U$ refer to the encryption keys of $S$ and $U$, respectively.

During the enrollment processes of $V^2$, $U$ gives a biometric sample to the $C$.
$C$ takes that sample and creates feature vectors from the sample.
The feature vectors are then sectioned into blocks.
$C$ then creates a chaff block for each of the feature vector blocks in such a way that it is indistinguishable from the real feature vector blocks to anyone but a user having the same original features.
The chaff blocks are created by randomly taking a corresponding feature vector from a different person\cite{vv-wacv2012}.
As shown in Figure \ref{vv_enroll}, each block is then encrypted, first using $K_U$ and again using $K_S$.
The blocks are then sent to $S$ in such a way that $S$ knows of the pairs which blocks are real and which blocks are chaff.

\begin{figure}[b]
\centering
\resizebox{140mm}{!}{\includegraphics{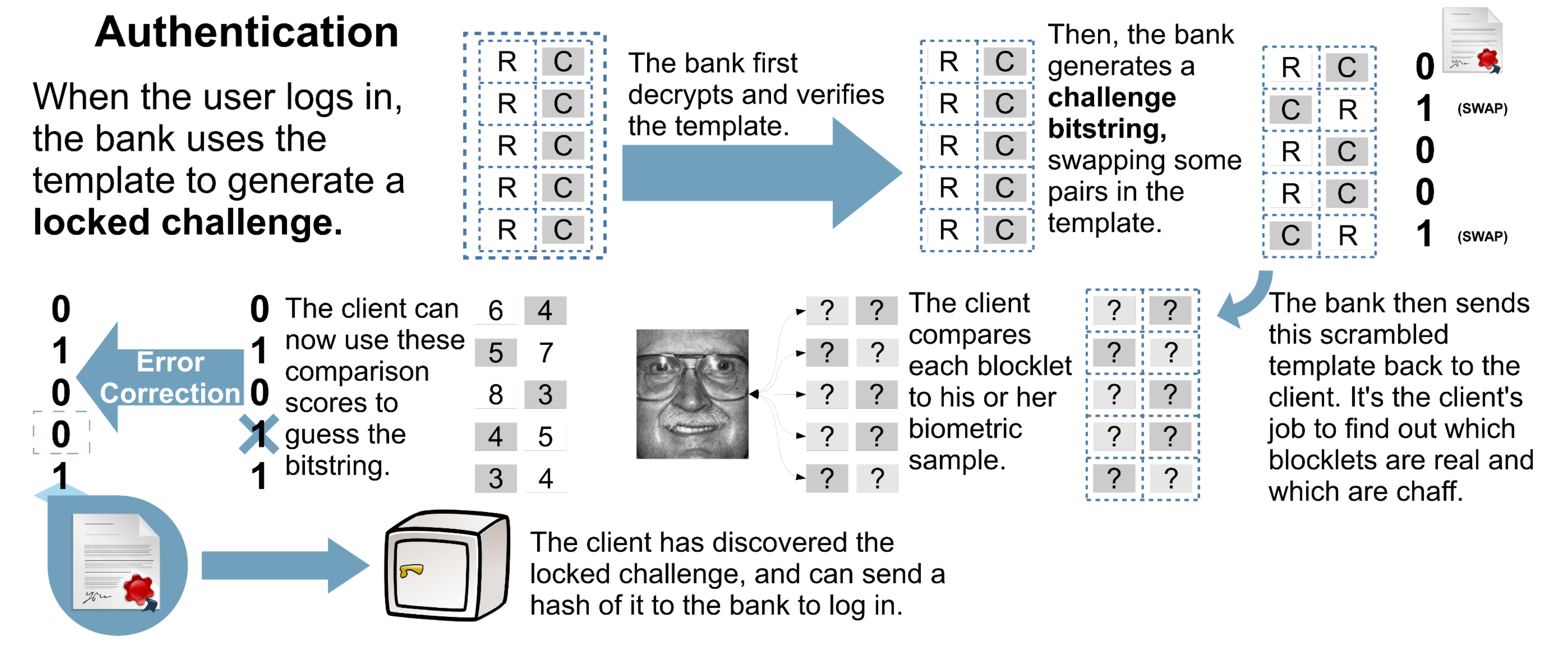}}
\caption{Visual example of \vv authentication/verification \cite{vv-wacv2012}.}
\label{vv_auth}
\end{figure}

As can be seen in Figure \ref{vv_auth}, when $U$ logs in during the verification process $S$ will send $U$ a locked challenge. 
For this, $S$ creates a random binary bitstring of $n$ bits such that roughly $n/2$ bits are 1's.
$S$ then sends $n$ block pairs to $C$ in an order that depends on the value of the binary bitstring; if the bit is a 0 the real block is sent first, if the bit is a 1 the chaff block is sent first.
$U$ then submits an image to $C$.
$C$ will then decrypt and analyze each block pair and respond with a 0 or 1 for each pair depending on which $C$ believes is real and which $C$ believes is chaff.

In this scenario, the blocks stored on $S$ are secure even if $S$ is compromised because $S$ does not have the encryption keys and, therefore, does not know the contents of the blocks.
The only information $S$ has is the knowledge of real vs chaff.
$S$ does not even concern itself with how the blocks are authenticated, just whether or not the responses are correct.

\subsection{Limitations}

This section describes the limitations of the \vv protocol.
One of the main limitations of \vv is that with the limited amount of information available in face and iris data.
\vv is not able to vary the data in the challenge-response process.
Also, because of the limited amount of data, there are only so many challenge-responses that can be generated.
The data limitations of \vv that are imposed by face and iris data are the same for both face and iris, so this explanation will refer only to face as shown in FIgures \ref{vv_enroll} and  \ref{vv_auth}.

As a part of the enrollment, the image is split into blocklets.
These blocklets are what get mixed with chaff.
Choosing between the real and the chaff blocklets is the basis for vaulted verification.
When sectioning the image into blocklets, a large enough section must be chosen so differentiation can happen.
If the sections are two small the real and the chaff will blend together, defeating the purpose \vv serves.
Inversely, the larger the blocklets the smaller the number of challenge-response pairs.

\vv will provide security with privacy.
The limitation becomes notable  when certain pairs are consistently guessed correctly by an attacker, rendering them useless.
The more pairs an attacker can consistently guess correctly, the weaker the security of the algorithm becomes.
Because of the limited data, there is a limited number of paris that can be generated, which in turn limits the effective security.


\section{\vvv}
\label{sec:vvv}
This section will describe the general procedure of how \vvv will be utilized in the creation of this novel scheme.
\vvv improves upon \vv in multiple ways.
\vvv can expand the available data using different words and phrases.
If some models are compromised they can be discarded and other models generated using new words and phrases.
An additional way is a further generalization of \vv via its application using voice.
Another is the enhanced variability and security through the question selection process.
With \vvv  $S$ does not know what the questions or the answers are; only that for some encrypted question, there exists a correct response among the encrypted choices.
Also, \vvv will be easy to use and implement on a mobile platform because there is no need for specialized hardware beyond a microphone, which all mobile phones have.

To understand the \vvv protocol, let us also define the following terms and symbols. 
$P$ denotes the password.
$q$ denotes a single question and $Q$ a set of questions.
$r$ denotes a single response and $R$ a set of responses.
The symbol $=>$ denotes transmission of some form (speaking to, typing in, sending over the network).
The symbol $<=>$ denotes back and forth communication.
$M$ denotes a message that is either an input or transmission.
\subsection{Enrollment Process}
\label{sec:enroll}

\begin{figure}
\centering
\resizebox{120mm}{!}{\includegraphics{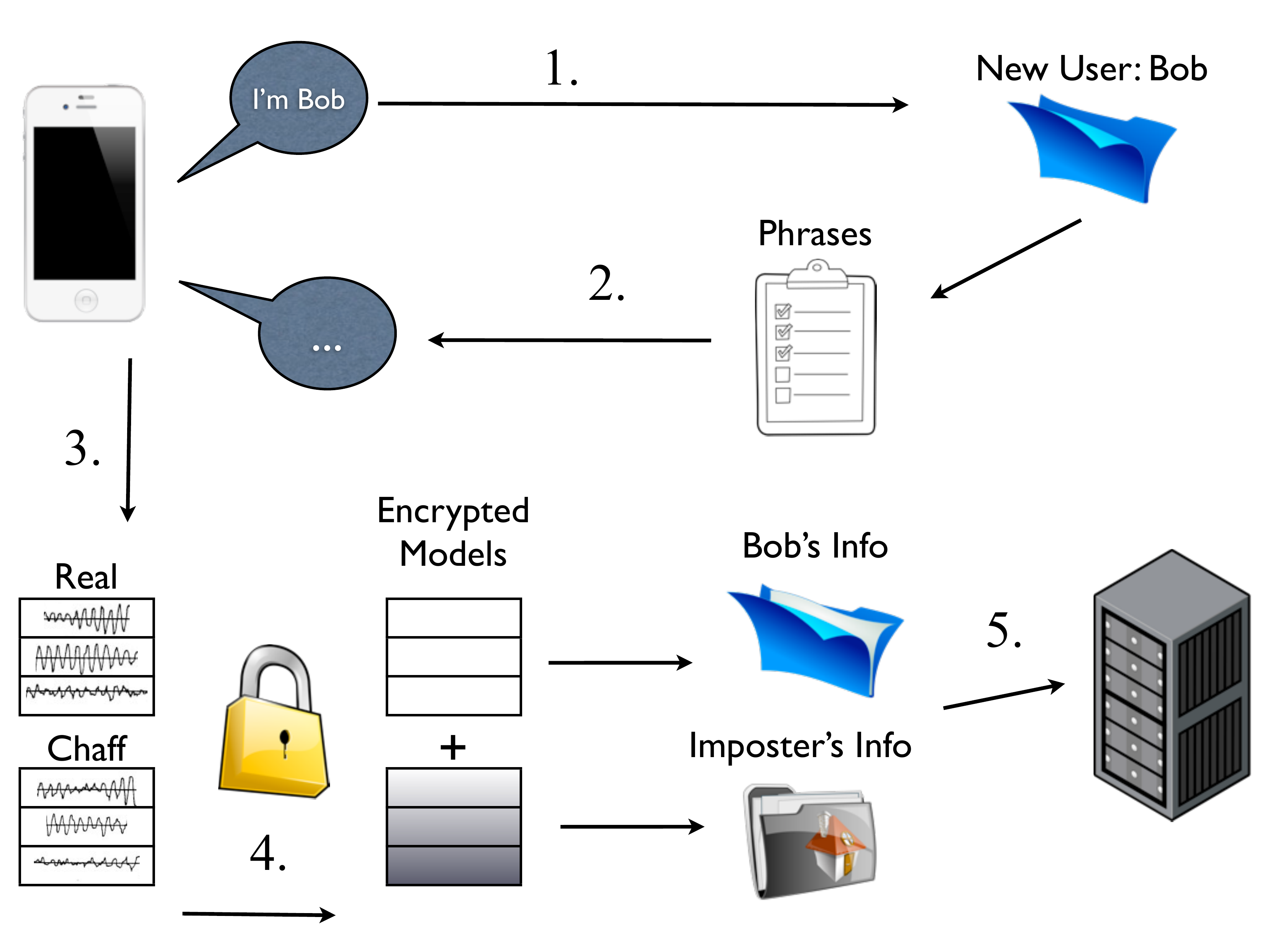}}
\caption{\vvv - Enrollment Process.}
\label{vvv_enroll}
\end{figure}

The first thing that must happen is enrollment.
A simplified version of the enrollment process is shown in Figure \ref{vvv_enroll}.
The steps mentioned in this section refer to the numbered arrows in Figure \ref{vvv_enroll}.
For layered security, during transmission $M$ will first be encrypted with private $K_U$ and again with public $K_S$.
$K_S$ is obtained when initial contact for enrollment is made from $C$ to $U$.
$U$ needs to give enough information so that they will be able to verify their identity at a later time.
The process involves $U$ supplying $S$ with some different information.

The first step, labeled step 1, is for $U$ to give $C$ identifying information, so $S$ can identify $U$ with their data.
$U$ tells $C$ their name and possibly other pieces of identification that $S$ will be able to use during the verification process.
$U$ also either inputs $P$ or a random $P$ is automatically generated and stored on $C$.
Next $C$ gives $S$ the identifying enrollment information.
This information is also encrypted $K_U$, so the only thing $S$ knows is an id and a hash.

\begin{equation}
U( id, P ) => C
\end{equation}

In step 2, $C$ asks $U$ to repeat a series of phrases/questions.
These phrases are what $S$ will use to challenge $C$ during verification.
The questions are designed to get the maximum amount of information out of the fewest number of phrases. 
The goal of the questions is to be able to generate a prescribed number of challenge-response pairs, thus having enough bits to give proper identity security.

\begin{equation}
S(q) = > C(q) => U
\end{equation}

As step 3 illustrates, $C$ will create adapted models from the responses.
The models created are GMMs.
Each GMM contains a number of distributions, $D$, and each of those distributions has a number of components, $C$, for a total number of components of $N = D*C$.
How these models are used and compared will be detailed in section \ref{sec:vv_process}.

The models and transcriptions of the responses are stored together.
Each of the model/transcription pairs are encrypted as a block of data.
Step 4 shows the models as they are first encrypted with $K_U$, and then again with $K_S$.
This will ensure that $S$ has no knowledge of what the responses/questions are.
The only thing $S$ is aware of is that for some encrypted question, there exists a correct response among the encrypted responses provided.

\begin{equation}
U( r ) => C
\end{equation}

$S$ then receives and stores the challenge/response pairs for future verification, illustrated in step 5.

\subsection{Verification Process}
\label{sec:vv_process}

\begin{figure}
\centering
\resizebox{120mm}{!}{\includegraphics{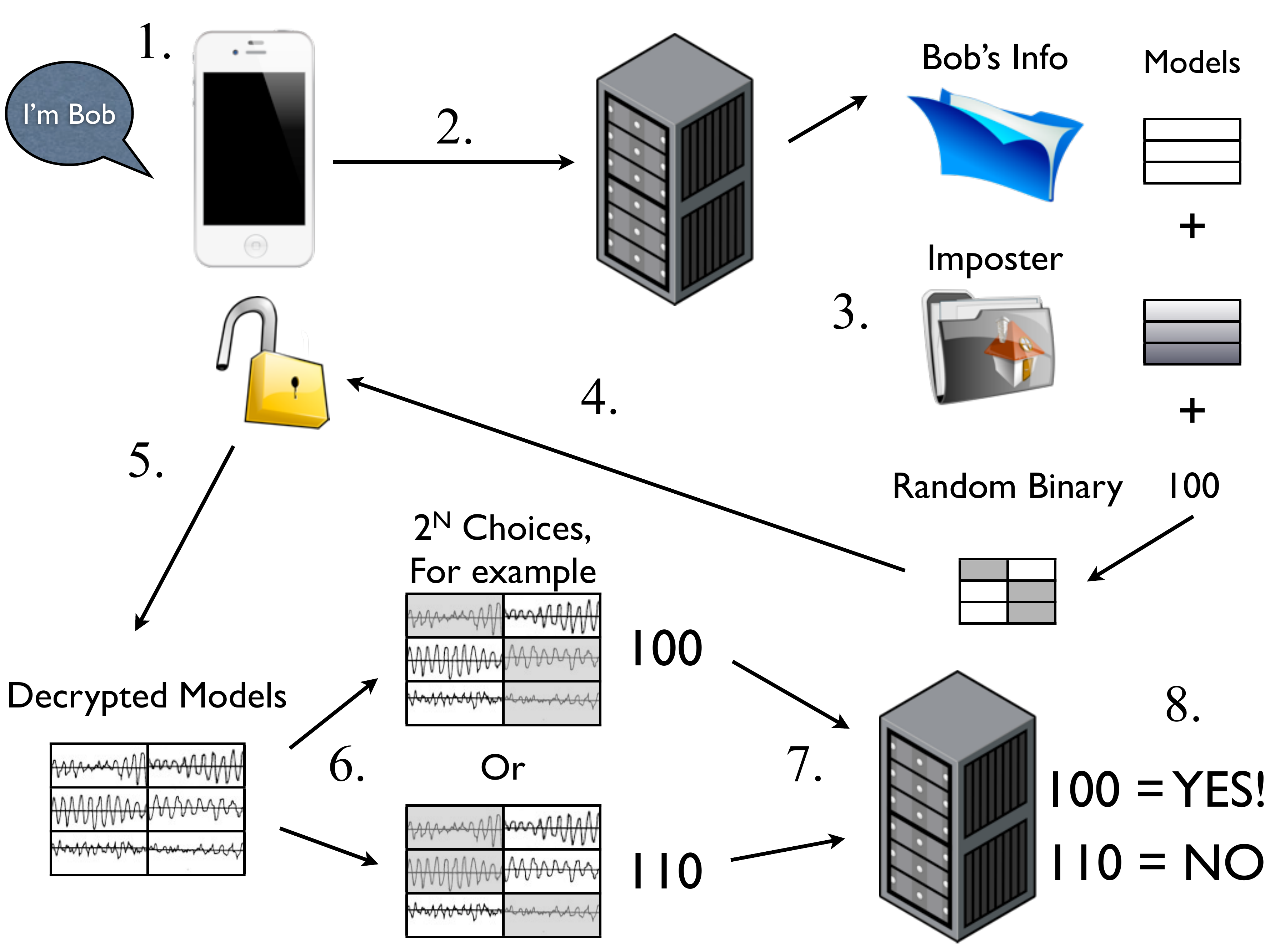}}
\caption{\vvv - Verification Process.}
\label{vvv_verify}
\end{figure}

The verification process is initiated when a user attempts to gain access to the system.
A simplified version of the verification processes is shown in Figure \ref{vvv_verify}.
All steps mentioned in this section refer to Figure \ref{vvv_verify}.
In this scheme, a user responds to questions asked by $C$.
This happens in steps as follows.

 First, $U$ tells $C$ that they would like to log in and access their data.
At this point, $C$ does not know who $U$ is or even who $U$ is claiming to be.
$U$ tells $C$ who they are by either entering in their name or speaking their name/id.
This is illustrated in step 1.

\begin{equation}
U(id) => C
\end{equation}

Then $C$ asks for the initial password/passphrase
While this happens, $C$ also sends a request to $S$ for $U$'s information, as illustrated in step 2.
$C$ uses $K_U$, derived from $P$, to decrypt the transmissions from $S$.
If $C$ does not have the correct $K_U$, there will be no way to get the rest of the information for the challenge responses, ending the process.
\begin{equation}
U( P ) => C, C(id) <=> S
\end{equation}

Once $C$ is able to decrypt the information packet, $M$, from $S$, $C$ compares the response to the information received from $U$.

\begin{equation}
K_U( M ) = id \ or \ 
K_U( M ) = 0
\end{equation}

If $C$ is able to decrypt $M$ using $K_U$, the challenge response process begins.
In the challenge response process, $S$ will send the $Q$ to $C$.
$Q$ is comprised of reordered pairs of $(q,r)$ and is appended with a nonce.
The pairs, as illustrated in step 3, have been reordered based on a random binary string that $S$ generates.

\begin{equation}
S( Q ) => C
\end{equation}

Step 4 illustrates $S$ sending encrypted $Q$ to $C$.
$Q$ is block encrypted with public $K_U$ before transmission.
This ensures that the blocks are different with every transmission; based on the random binary and the nonce.
In step 5 $C$ then decrypts and displays/asks $Q$ to $U$, one at a time, $q$.

\begin{equation}
C( q ) => U
\end{equation}
$U$ then responds to $q$.
The response depends on the type of question asked.
If the question is multiple choice, $U$ would speak the correct answer.
If the question is a passage to read, then $U$ simply reads what is asked.
\begin{equation}
U( r ) => C
\end{equation}

$C$ processes the response into a model.
A decision is made by comparing the response from $U$ to the options presented from $S$.
Step 6 illustrates an example of two choices that could result from comparing real and imposter/chaff models.
When comparing the similarity of two models, the probe, $pr$, and the gallery, $g$, either the variance of $pr$ or $g$ will be used depending on if the intention is to see how similar $pr$ is to $g$ or how similar $g$ is to $pr$.
Observe that for each gaussian mixture component there exists a mean, $\mu$, and a variance, $\sigma$.
In equation \ref{eq:zscore}, we are seeing how similar $pr$ is to $g$, which is to utilize the variance of $g$.
As shown in equation \ref{eq:totalscore}, a final score is the summation of the z-scores over the total number of components, $N$.

\begin{equation}
\label{eq:zscore}
Z_i = \frac{x_{pr} - \mu_{g}}{\sigma_{g}}
\end{equation}

\begin{equation}
\label{eq:totalscore}
Score = \sum_i^N{(Z_i)}
\end{equation}

Once the models are compared, $C$ will make a decision as to which of the models from $S$ is closest to the model generated from $U$.

Illustrated in step 7, $C$ will send a binary encoded response back to the server.
How this response will be encoded depends on the type of question.
If the question has two options, then 0|1 are appropriate.
For a multipart question, multiple bits may be required.
The process will repeat until stopping condition.
The stopping condition will generally be the required number of bits being reached.

After this exchange has finished, $C$ will have generated a large bitstring to send the server.
The bitstring will then be evaluated by the server.
If the score is above threshold, access will be granted, illustrated in step 8.

\begin{equation}
C (bitstring) => S
\end{equation}

\begin{equation}
S (yes\ or\ no) => C 
\end{equation}

\subsection{Relation to Prior work}

Comparing our work to the prior work as discussed in section \ref{related}, we can see that \vvv differs from the other techniques in a few different ways.
One of the ways our work is different is that in the enrollment phase; once the features are extracted and the model is generated, the process is over.
Also, in our technique the server will not access the raw data in any way and therefore, needs to have no knowledge of its composition.
The server has no knowledge of the data and cannot decrypt it. 
This is a novel and cryptographically stronger approach than simple  template hardening and obfuscation.

Because the server has no knowledge of the data and can not decrypt it, even if the server is fully compromised, the user's biometric data will be safe.
Even if the user's password were compromised, the approach uses text-dependent models and therefore, independent phrases can be revoked.

\section{Performance Evaluation}

The dataset selected for the tests is the MIT mobile device speaker verification corpus \cite{mit_data2006}.
This dataset was chosen because it appears in the work of Lopresti \cite{lopresti2011}, and is the state of the art benchmark against which to measure results.
The speech corpus is comprised on 48 speakers; 22 female and 26 male.
Short phrases, names and ice cream flavors, were recorded in 20 minute sessions.
Enrollment data was created in two separate sessions.
The imposter data was recorded in a separate session.
Each person has a dedicated imposter, meaning all imposter files for a user come from one person instead of each file being made from a different person's voice.

In order to perform the tests, the data needed to be separated.
The two enrollment sessions were split into a gallery set and a probe set.
At random, 60\% of the data was designated as gallery and the other 40\% as probe.
This ensures an appropriate balance between sessions.
The imposter data was used in its entirety as the imposter data.
In the tests, the data are further separated on a per-phrase basis; this way speech-dependent models could be created on which to run the tests.

The models created from the data are Gaussian Mixture Models, or GMMs.
GMMs are used because of their ability to model arbitrary data in a meaningful way.
The first step in modeling the voice is building the feature vectors.
The feature vectors are created by analyzing the audio spectrum.
Different groups \cite{zheng2001, murty2006}  have explained in greater detail how feature vectors are created, however, the general idea behind how the audio spectrum is broken up and turned into feature vectors can be explained in a fairly straightforward manner.
The audio spectrum is divided into slightly overlapping frequency blocks and then a transform of some type, typically a DFT, is used to transform each of those blocks into cepstrum coefficients.
The type of cepstrum coefficients used is referred to as Mel Frequency Cepstrum Coefficients (MFCCs).
The MFCCs are what make up the feature vectors.
The feature vectors are used to describe a given audio sample in a meaningful and quantifiable way.

Once the feature vectors are created, a model can be generated.
The model that is created can model almost anything about the audio sample depending on how the samples are grouped, and how the feature vectors are created.
Gaussian distributions have been found to be extremely useful in modeling the created feature vectors.
For a deeper understanding of gaussian distributions and why they are useful for modeling based on feature vectors refer to \cite{rao2010}.

The number of distributions chosen is dependent of the amount of data that are being modeled.
The larger the number of distributions used, the finer the granularity of the model. 
If only a small amount of data is being modeled then there will not be enough data to necessitate a larger number of distributions.

When looking into how to compare the different models many options were found that could have been suitable.
The option finally chosen was z-scores, as shown in Equation \ref{eq:zscore}.
Z-scores, also known as standard scores, were chosen for a number of reasons, the best of which are the simplicity of computation and the relative proximity of the components of the formula to the model data.

Performance is measured in terms of the false accept rate (FAR) and the false reject rate (FRR).
FAR measures the percentage of people that are allowed access to the system what they are posing as someone else.
FRR measures the percentage of people who are falsely denied access to the system.
Where these values meet when plotted against each other is called the equal error rate (EER).
It is crucial to mention that the EER in \vvv only applies to security after the attacker has cracked/compromised the keys ($K_U$ and $K_S$).
Before these keys are compromised, an attacker would get nothing.

\begin{figure}
\centering
\resizebox{100mm}{!}{\includegraphics{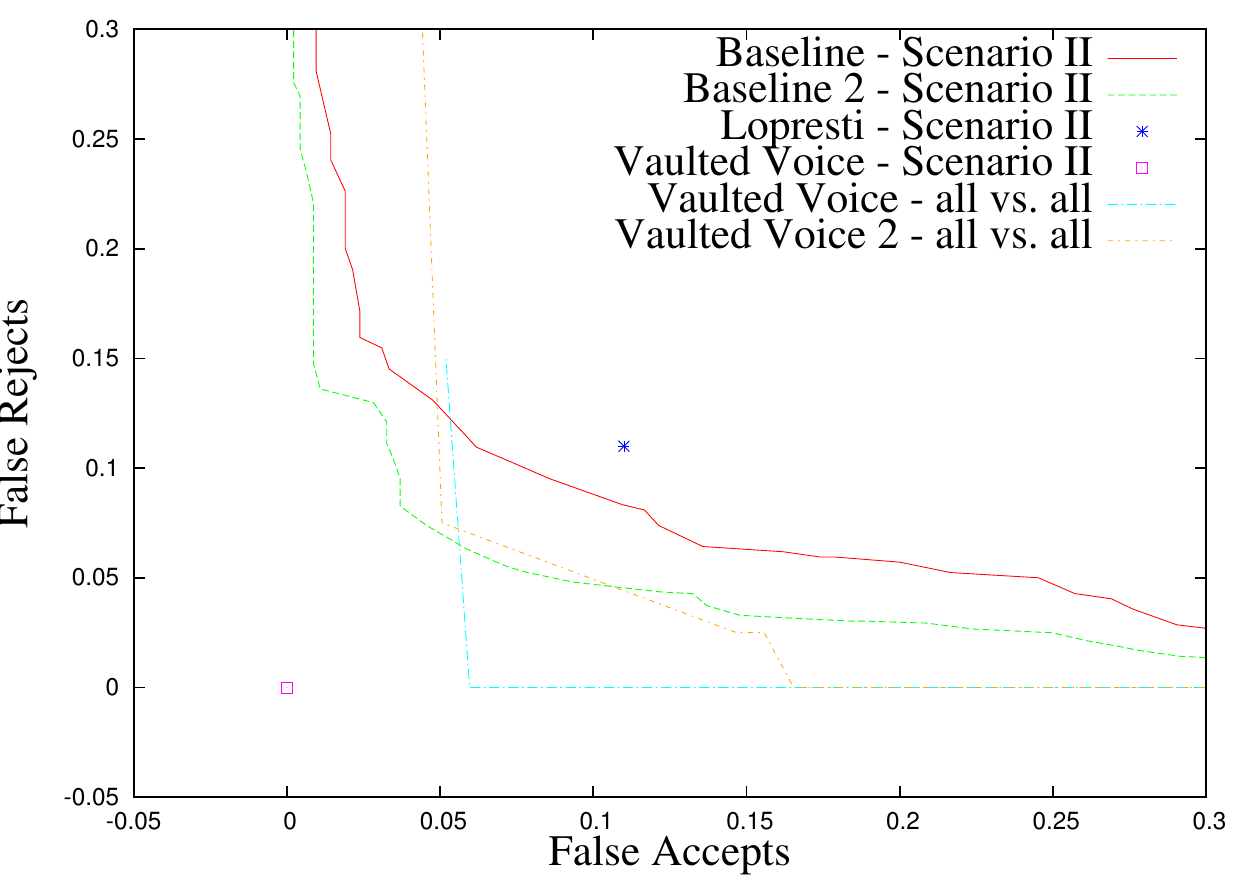}}
\caption{ROC plots from a baseline algorithms on Scenario-II as well as \vvv on a larger ``all-vs-all'' scenario with a larger imposter set. Shown as points are  \vvv  and Lopresti's work on Scenario II, with an ERR of 0\% and 11\% \cite{lopresti2011}, respectively. }
\label{roc_spie}
\end{figure}

In these experiments, we are assuming all keys have been compromised and examining the amount of security \vvv adds on top of the encryption. 
This is what Lopresti\cite{lopresti2011} calls Scenario II, for which he achieved 11\% EER. 
A test was run to make a direct comparison to the work done by Lopresti\cite{lopresti2011, lopresti2012}.
This test also assumed that all keys had been compromised, as was done in Scenario II of Lopresti's work.
This test did not cover the scenario in which the passwords and keys were not compromised because if the data is not decrypted no valid answer can be supplied by the attacker. 
This test was implemented by comparing the proximity of $g$ to $pr$ and $pr_i$ for each given phrase for an individual. 
The $pr$ and $pr_i$ pairs were presented in a random order and $g$ had to decided which it was closest to and respond accordingly.
In this test there existed only two choices per phrase, so each response was binary.
In this test an EER of 0\% was generated.
Plotting a 0\% EER does not illustrate anything compelling, so it was plotted as a single point on Figure \ref{roc_spie}.
It is necessary to note, however, that this test directly compares to the test in which Lopresti reported 11\%, but the ERR of 0\% is unrealistic because of the small number of impostors.

To provide a stronger experimental validation, we implemented a new baseline algorithm, which is not privacy/security enhanced, and then expanded the testing to a larger test set. 
Figure \ref{roc_spie} shows the ROC curves for these test results.
There are results from the two baseline tests and results from the two \vvv expanded tests.
The baseline tests were created by comparing the scores between $pr$ and $g$ in the same manner as \vvv.
However, in the baseline tests only raw scores were recorded, removing the inherent pairwise thresholding that is added by \vvv.

To generate the data for the first baseline curve, each person in the gallery was tested against $pr$ and their dedicated imposter, $pr_i$.
In these tests, the variance of $g$ is used.
This gives an approximate equal error rate of 8\%.
A second baseline was also generated (labeled Baseline 2).
This baseline was created to show a similar separation of data exists when using the variance of $pr$ as opposed to using the variance of $g$. 
The EER of this baseline test is 6\%. Both of these results show an improvement over the state of the art benchmark of 11\%. We note, however, the baseline has no privacy enhancements.

\vvv performed significantly better than the baseline because \vvv has an inherent pairwise thresholding step which allows it to make every decision on a case by case bases.
For demonstration purposes, we give the following scenario of two phrases.
In this scenario, let us assume that a score can range from 0 to 10.
For the first phrase, $ph_0$, $g$ generated a score of 5 for $pr_0$ and a score of 3 for $pr_{0i}$.
For the second phrase, $ph_1$, $g$ generated a score of 9 for $pr_1$ and 6 for $pr_{1i}$.
According to the baseline scores, no clear threshold exists that would separate probes and their impostors.
However, because \vvv thresholds on a case by case bases, it is able to identify and discern between the probes and impostors for both the first and second phrases.

To better estimate the EER of the \vvv protocol, a second set of tests were performed.
These tests are labeled in Figure \ref{roc_spie} as all vs all. 
In the first all vs all test, labeled Voice Verification -all vs all, we tested all people in the gallery against all $pr$ and $pr_i$ pairs.
By doing so, we have expanded the negative matches to more thoroughly examine the false match rate of the algorithm.
In keeping with the tests performed to generate a baseline, a second all vs all test was performed.
This test, labeled Voice Verification 2 - all vs all, was done using the variance of the probe and imposter against the gallery data.
As Figure \ref{roc_spie} shows, these two tests have an EER of approximately 6\%.

\section{Security Analysis}

How is \vvv able to solve the security issues mentioned in the Introduction?
As with most secure biometric systems, this protocol contains many layers of security.
As mentioned previously, the base security of this protocol is similar to the work done by Wilber, et al \cite{vv-wacv2012, vv-cvpr2012}.


After enrollment, the client never sends out non-encrypted data, the matching is done on the client after decryption with  the user supplied password.  From enrollment to verification it is assumed that the communication is happening over a secure encryption protocol, SSL or TLS for example, and an attacker cannot ease drop on the communication.  If the encryption is maintained, the system is secure, so we consider various levels of compromise. 

If an attack is somehow able to break the encryption on the communication and try to impersonate $S$ and stage a man in the middle attack, the attacker would not be able to gain any additional information about the biometric to give them access.
This is because even if the attacker is able to see the data from $C$ to $S$, the data is still encrypted with $K_U$, as illustrated in step 4 of Figure \ref{vvv_enroll}.

If an attacker attempted a man in the middle attack by recording the encrypted and randomly ordered data pairs and responses, this would also prove pointless.
This is because $Q$ is randomly ordered, has a nonce added, and then is block-encrypted with $K_S$ for every session, making the transmitted blocks different for every session.
Thus any reordering would fail to decrypt and no information is gained by examining it since the recorded information would be different for the next session. With this, there is, effectively, no man in the middle.

If the device was lost, it provides little information as it just has an encrypted block of data that the client cannot decode. 
If an attacker were to obtain both the server's private key and the user's device, the attacker might be able to authenticate on that particular server.
This is because they would be able to decrypt the models and, through trial and error, return the correct response to the challenges via examining the differences when models are swapped.
However, because this is a phrase based scheme, this would not allow an attacker access the user's data on a different server.
The reasons are two fold.
The first is because the models are encrypted using the user's private key.
Without being able to access the models the attacker would not be able to discern any information about it.
The second is that without recordings of the correct phrases, the attacker would not have the necessary information to discern between real and chaff data.    If the device is lost,  the enrolled model can and should be revoked changing the encryption keys and phrases.

The worse case scenario for this protocol would be for the attacker to have access to $K_S$ and $K_U$.
The attacker would then be able to authenticate as $U$ on $S$, but would still not know which of the $N$ blocks are real and which are chaff.
On average, an attacker with  all the keys must still make $2^N$ guesses to correctly identify the biometrics.
At 56 blocks, that is $2^{56}$ attempts; not too difficult for computers today.
If the number of blocks were increased to 256, with current computational models, this is sufficiently secure.  And again, the user can just revoke and reissue with new phrases and keys. 

The lists used in the MIT dataset \cite{mit_data2006} had one dedicated imposter for each phrase.
Each phrase in the data set was repeated 6 to 10 times per person.
There were between 8 and 12 phrases spoken per person.
The tests performed were using per-phrase comparisons and where, therefore, limited to 8 to 12 bits.
That gives a probability of between $ 2^{-8}$ and $2^{-12}$ for an attacker with no foreknowledge, using a random chance brute force attack, to gain access.
Even with this, the error rate works out to between 0.39\% and 0.02\%, which is better than 11\% reported by Lopresti \cite{lopresti2011} under the same conditions, scenario II.
Our results are an improvement over straight biometrics, but we want better security.

To increase the security, we would need to increase the number of bits.
One of the ways to do this would be to use multiple impostors instead of the one dedicated imposter.
This would increase the number of bits based on the number of options.
For example, if there were 3 impostors per question instead of 1,  there would be 4 options from which to chose.
The probability for each question would be $2^{-2}$, effectively doubling the number of bits per question.
For the same 12 phrases, there would then be a probability of $2^{-24}$ that an attacker could authenticate as $U$ and gain access.

Again, it is necessary to note that this security analysis of \vvv is speaking in terms of how much security is added on top of encryption.
\vvv would be providing  $K$-bits  of ``identity verification" security in addition to the $N$-bits of security offered by encryption. 
Since the equal error rate provides a lower bound on identification accuracy, because of the associated biometric dictionary attack, the identity security offered by \vvv is already much better than the prior state of  the art.

\section{Conclusion and Future Work}

To conclude, with this work we have developed a verification protocol that addresses the instabilities inherent in voice in a way that still preserves privacy.
We described a novel protocol that allows a user to authenticate using voice  on a mobile/remote device without compromising their privacy.
We showed that this protocol, dubbed \vvv (or $V^3$), is able to achieve a 0\% equal error rate in on the same  tests where Lopresti \cite{lopresti2011} achieved 11\%.  The new protocol  achieves 6\% ERR on larger scale test. 

In future work, we will extend the challenge response protocol to further demonstrate the abilities of this protocol with voice.  We will look into expanding the challenge response protocol, improving security and robustness, by examining the types of questions than can be asked.

One of the ways this will be accomplished is by mixing the text-dependent work done so far with text-independent modeling; allowing us to extend beyond basic words and phrases.
Adding speech detection to the system also increases the number of bits.
If, during the verification process, the system is also able to determine what the spoken phrase is,  more bits would be added.
For example, the work done by Just \cite{just2009} showed that it could take as many as $2^{48} $ answers to correctly guess a word.
If asked an open-ended question, the number of bits would be increased significantly.
In addition to questions, we could also shuffle the order of phrases to increase the number of bits.

Another aspect we will look into is using location based questions and answers to decide which questions are asked. 
This will also help to defend against replay attacks.
We also look to extend the security requirements of biometric systems by examining the ability and implications of detecting verification attempts made under duress.

\acknowledgments
We thank Michael Wilber and Brian Heflin for their contributions to Figures \ref{vv_enroll} and \ref{vv_auth}.


\bibliography{report}   
\bibliographystyle{spiebib}   

\end{document}